\documentclass[twoside,twocolumn,9pt]{article}
\usepackage{extsizes}
\usepackage[super,sort&compress,comma]{natbib}
\usepackage[version=3]{mhchem}
\usepackage[left=1.5cm, right=1.5cm, top=1.785cm, bottom=2.0cm]{geometry}
\usepackage{balance}
\usepackage{mathptmx}
\usepackage{sectsty}
\usepackage{graphicx}
\usepackage{lastpage}
\usepackage[format=plain,justification=justified,singlelinecheck=false,font={stretch=1.125,small,sf},labelfont=bf,labelsep=space]{caption}
\usepackage{float}
\usepackage{fancyhdr}
\usepackage{fnpos}
\usepackage[english]{babel}
\addto{\captionsenglish}{%
  
}
\usepackage{bm}
\usepackage{array}
\usepackage{droidsans}
\usepackage{charter}
\usepackage[T1]{fontenc}
\usepackage[usenames,dvipsnames]{xcolor}
\usepackage{setspace}
\usepackage[compact]{titlesec}
\usepackage{hyperref}

\usepackage{epstopdf}

\definecolor{cream}{RGB}{222,217,201}

\begin{document}

\pagestyle{fancy}
\thispagestyle{plain}
\fancypagestyle{plain}{
\renewcommand{\headrulewidth}{0pt}
}

\makeFNbottom
\makeatletter
\renewcommand\LARGE{\@setfontsize\LARGE{15pt}{17}}
\renewcommand\Large{\@setfontsize\Large{12pt}{14}}
\renewcommand\large{\@setfontsize\large{10pt}{12}}
\renewcommand\footnotesize{\@setfontsize\footnotesize{7pt}{10}}
\makeatother

\renewcommand{\thefootnote}{\fnsymbol{footnote}}
\renewcommand\footnoterule{\vspace*{1pt}%
\color{cream}\hrule width 3.5in height 0.4pt \color{black}\vspace*{5pt}}
\setcounter{secnumdepth}{5}

\makeatletter
\renewcommand\@biblabel[1]{#1}
\renewcommand\@makefntext[1]%
{\noindent\makebox[0pt][r]{\@thefnmark\,}#1}
\makeatother
\renewcommand{\figurename}{\small{Fig.}~}
\sectionfont{\sffamily\Large}
\subsectionfont{\normalsize}
\subsubsectionfont{\bf}
\setstretch{1.125} 
\setlength{\skip\footins}{0.8cm}
\setlength{\footnotesep}{0.25cm}
\setlength{\jot}{10pt}
\titlespacing*{\section}{0pt}{4pt}{4pt}
\titlespacing*{\subsection}{0pt}{15pt}{1pt}

\fancyfoot[LO,RE]{\vspace{-7.1pt}\includegraphics[height=9pt]{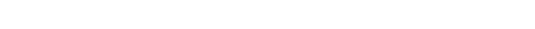}}
\fancyfoot[CO]{\vspace{-7.1pt}\hspace{13.2cm}\includegraphics{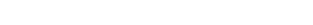}}
\fancyfoot[CE]{\vspace{-7.2pt}\hspace{-14.2cm}\includegraphics{head_foot/RF}}
\fancyfoot[RO]{\footnotesize{\sffamily{1--\pageref{LastPage} ~\textbar  \hspace{2pt}\thepage}}}
\fancyfoot[LE]{\footnotesize{\sffamily{\thepage~\textbar\hspace{3.45cm} 1--\pageref{LastPage}}}}
\fancyhead{}
\renewcommand{\headrulewidth}{0pt}
\renewcommand{\footrulewidth}{0pt}
\setlength{\arrayrulewidth}{1pt}
\setlength{\columnsep}{6.5mm}
\setlength\bibsep{1pt}

\makeatletter
\newlength{\figrulesep}
\setlength{\figrulesep}{0.5\textfloatsep}

\newcommand{\topfigrule}{\vspace*{-1pt}%
\noindent{\color{cream}\rule[-\figrulesep]{\columnwidth}{1.5pt}} }

\newcommand{\botfigrule}{\vspace*{-2pt}%
\noindent{\color{cream}\rule[\figrulesep]{\columnwidth}{1.5pt}} }

\newcommand{\dblfigrule}{\vspace*{-1pt}%
\noindent{\color{cream}\rule[-\figrulesep]{\textwidth}{1.5pt}} }

\makeatother

\twocolumn[
  \begin{@twocolumnfalse}
{\includegraphics[height=30pt]{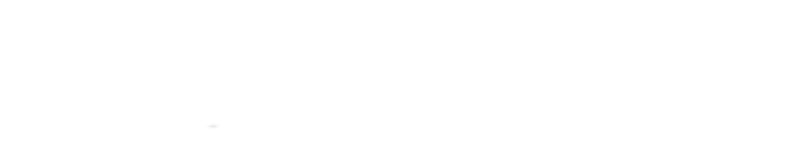}\hfill\raisebox{0pt}[0pt][0pt]{\includegraphics[height=55pt]{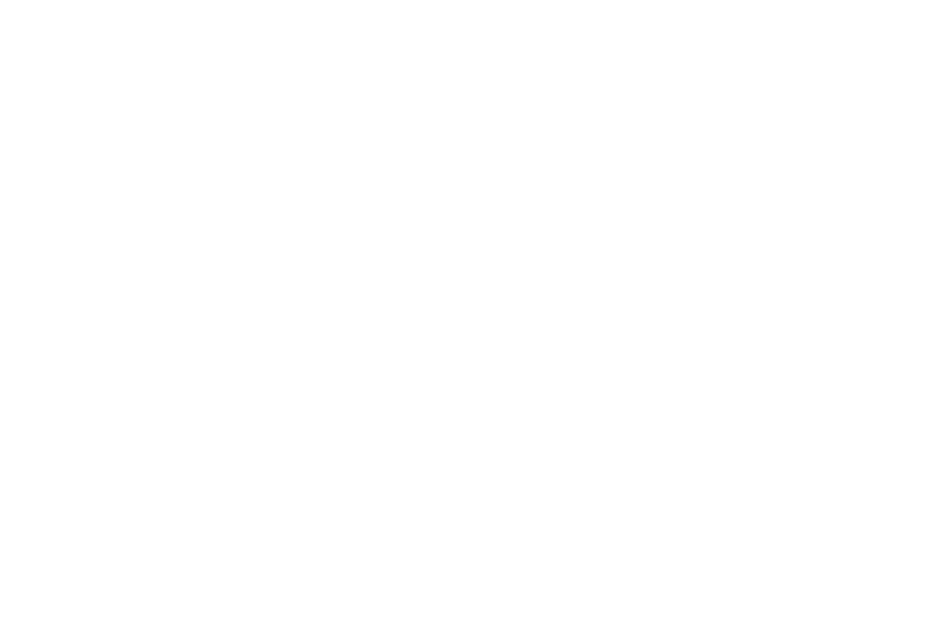}}\\[1ex]
\includegraphics[width=18.5cm]{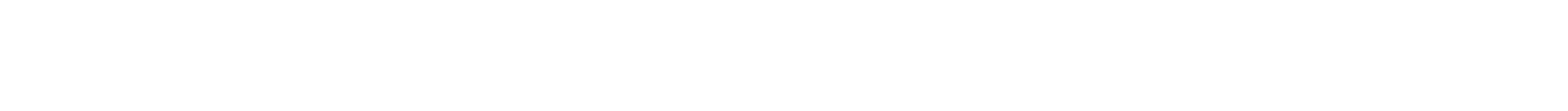}}\par
\vspace{1em}
\sffamily
\begin{tabular}{m{4.5cm} p{13.5cm} }

\includegraphics{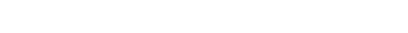} & \noindent\LARGE{\textbf{Three-dimensional compaction of soft granular packings}} \\
\vspace{0.3cm} & \vspace{0.3cm} \\

 & \noindent\large{Manuel C\'ardenas-Barrantes,\textit{$^{a,b}$} David Cantor,\textit{$^{c}$} Jonathan Bar\'es,\textit{$^{a}$} Mathieu Renouf,\textit{$^{a,b}$} and Emilien Az\'ema \textit{$^{a,b,d}$}} \\

\includegraphics{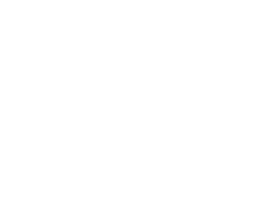} & \noindent\normalsize{
This paper analyzes the compaction behavior of assemblies composed of soft (elastic) spherical particles beyond the jammed state, using three-dimensional non-smooth contact dynamic simulations.
The assemblies of particles are characterized using the evolution of the packing fraction, the coordination number, and the von Misses stress distribution within the particles as the confining stress increases.
The packing fraction increases and tends toward a maximum value close to $1$, and the mean coordination number increases as a square root of the packing fraction.
As the confining stress increases, a transition is observed from a granular-like material with exponential tails of the shear stress distributions to a continuous-like material characterized by Gaussian-like distributions of the shear stresses.
We develop an equation that describes the evolution of the packing fraction as a function of the applied pressure.
This equation, based on the micromechanical expression of the granular stress tensor, the limit of the Hertz contact law for small deformation, and the power-law relation between the packing fraction and the coordination of the particles, provides good predictions from the jamming point up to very high densities without the need of tuning any parameters.
} \\

\end{tabular}

 \end{@twocolumnfalse} \vspace{0.6cm}

  ]

\renewcommand*\rmdefault{bch}\normalfont\upshape
\rmfamily
\section*{}
\vspace{-1cm}


\footnotetext{\textit{$^{a}$~LMGC, Universit\'e de Montpellier, CNRS, Montpellier, France. \\
manuel-antonio.cardenas-barrantes@umontpellier.fr,\\ Jonathan.bares@umontpellier.fr,\\mathieu.renouf@umontpellier.fr,\\emilien.azema@umontpellier.fr}}
 \footnotetext{\textit{$^{b}$~Laboratoire de Micromécanique et Intégrité des Structures (MIST), UM, CNRS, IRSN, France.}}
\footnotetext{\textit{$^{c}$~Department of Civil, Geological and Mining Engineering, Polytechnique, 2500, chemin de Polytechnique, Montréal, Qu\'ebec, Canada.
 david.cantor@polymtl.ca}}
\footnotetext{\textit{$^{d}$~Institut Universitaire de France (IUF), Paris, France.}}

\section{Introduction}

The importance of understanding the physics behind the compaction of granular systems made of soft particles lies in the numerous natural phenomena and human activities that deal with such kind of materials.
They are present from constitutive biological cells, foams, and suspensions \cite{Katgert2010_Jamming, Chelin2013_Simulation, Dijksman2017, Katgert2019_The} to powder compaction, pharmaceutical industries, and food activities \cite{Heckel1961, Montes2010, Parilak2017, Montes2018}.
In some civil engineering construction, mixing coarse grains with rubber residues exhibit surprising properties such as better stress relaxation  \cite{Indrarantna2019_Use, Khatami2019_The, Anastasiadis2012_Small, Platzer2018} or better foundation damping \cite{Mashiri2015_Shear, Senetakis2012_Dynamic, Kianoosh_2021}.

In particular, far beyond the jamming point, the compaction behavior of soft granular materials is a vast and still open subject, with notable experimental, numerical, and theoretical challenges.
Among these challenges, a three-dimensional characterization, by a realistic model that would consider both the change in grain shape and the assemblies' multi-contact aspects, remains poorly studied.

In experiments, an underlying difficulty is to track the change in particle shape while detecting the making of new contacts.
Photoelasticimetry \cite{ Howell1999, KDaniels2017, Zadeh_2019} or inverse problem method coupled with Digital Image Correlation (DIC) \cite{HURLEY2014154, Marteau2007} is the most used experimental technic for analyzing hard particles' two and three-dimensional behavior.
In particular, the DIC method, which has been very recently extended to analyze two-dimensional soft particle assemblies far beyond the jammed state \cite{Vu2019, Vu2019a}, directly quantifies the deformation field inside the particles and characterizes the deformation mechanisms.
However, for high packing densities, the image resolution may sharply limit the tracking of the grains and the detection of contacts between highly deformed particles, which is crucial for three-dimensional geometries.
Furthermore, in three dimensions, it is not always possible to use an optical approach to measure local properties of the particles, and tomography reconstructions may be necessary, but technically laborious \cite{Dijksman2017,MTiel2017,bares2020_Transparent, Ando2020}.

Concerning numerical modeling, the discrete element method (DEM), coupled with a complementary approach such as the Bonded-Particle Method (BPM) or the Finite-Element Method (FEM), is a suitable framework to simulate and analyze the compaction of soft particles assemblies.
In DEM-BPM, deformable particles are seen as aggregates of rigid particles interacting via elastic bonds \cite{dostaNumericalInvestigationCompaction2017, Nezamabadi_2017, Asadi2018_Discrete}.
This approach is relatively straightforward and allows one to simulate 3D packing composed of a large number of aggregates.
However, a major drawback is that the deformable particles often present a plastic behavior at large strain, and their characterization can be complex depending on the imposed numerical parameters (e.g., size of primary particles or interaction laws) \cite{Azema2018, voMechanicalStrengthWet2018a}.
In contrast, DEM-FEM strategies have the advantage of being closely representative both in terms of geometry and bulk properties of the particles. The price to pay is that these simulations are computationally expensive.
DEM-FEM methods can be classified into two classes. The Multi-Particle Finite Element Method (MPFEM) \cite{Munjiza_2004}, in which regularized contact interactions are used, and the Non-Smooth Contact Dynamic Methods (NSCD) \cite{Moreau1994_Some, Jean1999}, which uses non-regularized contacts laws.
To our best knowledge, the first MPFEM simulations applied to the compaction of deformable disks were performed in the 2000s \cite{Gethin_2002, Procopio2005}, and the first 3D compaction simulations appear a few years later \cite{Shmidt_2010, HARTHONG2012784, ABDELMOULA2017142, ZOU2020297, PENG2021478}.
The first applications of the NSCD to the compaction of soft grains assembly are reported in recent works by Vu et al. \cite{Vu2019, Vu2020_compaction, Vu2021_Effects} for two-dimensional hyper-elastic disks.
In practice, an inherent difficulty to DEM-FEM methods that explains the small number of studies, particularly in 3D, is the high computational cost, limiting the number of particles that can be simulated \cite{HARTHONG2012784, ABDELMOULA2017142, PENG2021478}.

Finally, a lack in the description of the microstructural phenomena during the compression limits the development of theoretical models upon the compaction of soft grain assemblies (i.e.,  a relation between the applied pressure $P$ and the evolution of the packing faction $\phi$).
As reviewed in literature \cite{Heckel1961, Panelli2001, Comoglu2007, Denny2002, Popescu2018_Compaction, Montes2018, Platzer2018, nezamabadiModellingCompactionPlastic2021}, many compaction equations have been proposed during the last decades.
In general, existing models are based on macroscopic assumptions, and, thus, fitting parameters are required to adjust each expression to the data.
Among these equations, the most used is the one proposed by Heckel \cite{Heckel1961} and later improved by Secondi \cite{Secondi2002_Modelling}. It states that $P \propto \ln(\phi_{max}-\phi)$, where $\phi_{max}$ is the maximum packing fraction that the assembly can reach.
Carroll and Kim justified this equation by an analogy between the corresponding loss of void space and the collapse of a cavity within an elastic medium under isotropic compression \cite{Carroll1984, Kim1987}.
Depending on the authors, different interpretations have been provided to the fitting parameters, which are supposed to represent either a characteristic pressure, a hardening parameter, or is linked to the assembly's plasticity \cite{Platzer2018, Montes2018}.
Only recently, Cantor et al. \cite{Cantor2020_Compaction} and Cardenas-Barrantes et al. \cite{Cardenas_pentagons_2021} set up a systematic micro-mechanical
approach to study the compaction of soft granular assemblies.
By applying this framework to two-dimensional systems modeled with NSCD simulations, new compaction laws entirely determined through the evolution of the connectivity of the particles and the contact properties were presented.

This article presents a three-dimensional numerical and theoretical analysis of the compaction of assemblies composed of highly deformable (elastic) spherical particles using the Non-Smooth Contact Dynamics Method.
We are interested in the compaction evolution as a function of the applied stress from the jammed state to a packing fraction close to unity. 

As mentioned before, similar studies were recently performed in 2D \cite{Cantor2020_Compaction, Cardenas_pentagons_2021}.
The transition from 2D to 3D requires additional numerical and technical efforts to manage the particles' deformation correctly.
Furthermore, as we will see, the resulting compaction equation differs from those previously established in 2D using the same micromechanical framework.
This will be understood from the approximation of the Hertz's contact law in the small deformation regime, where the contact force between two deformable particles shifts from a linear dependence with the contact deflection in 2D to a power-law dependence in 3D.

The paper is organized as follows. In Section \ref{Numerical_method}, we introduce the numerical framework used in the simulations.
The numerical results and the theoretical model of the 3D compaction curves are discussed in Sec.\ref{Sec_results}.
Section \ref{Stress_evolution} deals with the evolution of the packing fraction and particle connectivity beyond the jamming point as a function of the applied stress.
In Sec. \ref{Theory}, we present the micro-structural elements behind the evolution of the packing fraction and the corresponding resulting 3D equation.
In Section \ref{Local_stress}, a more refined description of the particle stresses is presented within the limit of the representativeness of the considered samples.
Finally, some conclusions are discussed in Section \ref{Conclu_section}.

\section{Numerical Approach}
\label{Numerical_method}

\subsection{The Non-Smooth Contact Dynamic Method (NSCD)}
The simulations are performed using the Non-Smooth Contact Dynamics (NSCD), a method developed by Moreau and Jean \cite{Moreau1994_Some, Jean1999, Dubois2018}.
The NSCD extends the Contact Dynamic (CD) method \cite{Moreau1994_Some} to deformable bodies through a finite element approach (FEM).

The CD method is based on an implicit time integration of the equations of motion and non-regularized contact laws.
These contact laws set the non-penetrability and friction behavior between the particles.
No elastic repulsive potentials and no smoothing of the Coulomb friction law are needed to determine the contact forces.

Therefore, the unknown variables, i.e., particle velocities and contact forces, are simultaneously solved via a nonlinear Gauss-Seidel scheme.
Considering deformable bodies (in the sense of continuous mechanics) is natural with CD, although technically very complex to implement.
In this case, the bodies are discretized via finite elements, so the degrees of freedom - the coordinates of the nodes - and contact interactions are resolved simultaneously.

We used an implementation of the three-dimensional Non-Smooth Contact Dynamics Method available on the open-source software LMGC90, capable of modeling a collection of deformable or non-deformable particles of various shapes, behaviors and interactions \cite{Dubois2006}.

\subsection{Packing composed of 3D elastic particles}
\begin{figure}
\centering
\hspace{26pt} \includegraphics[width=0.7\linewidth]{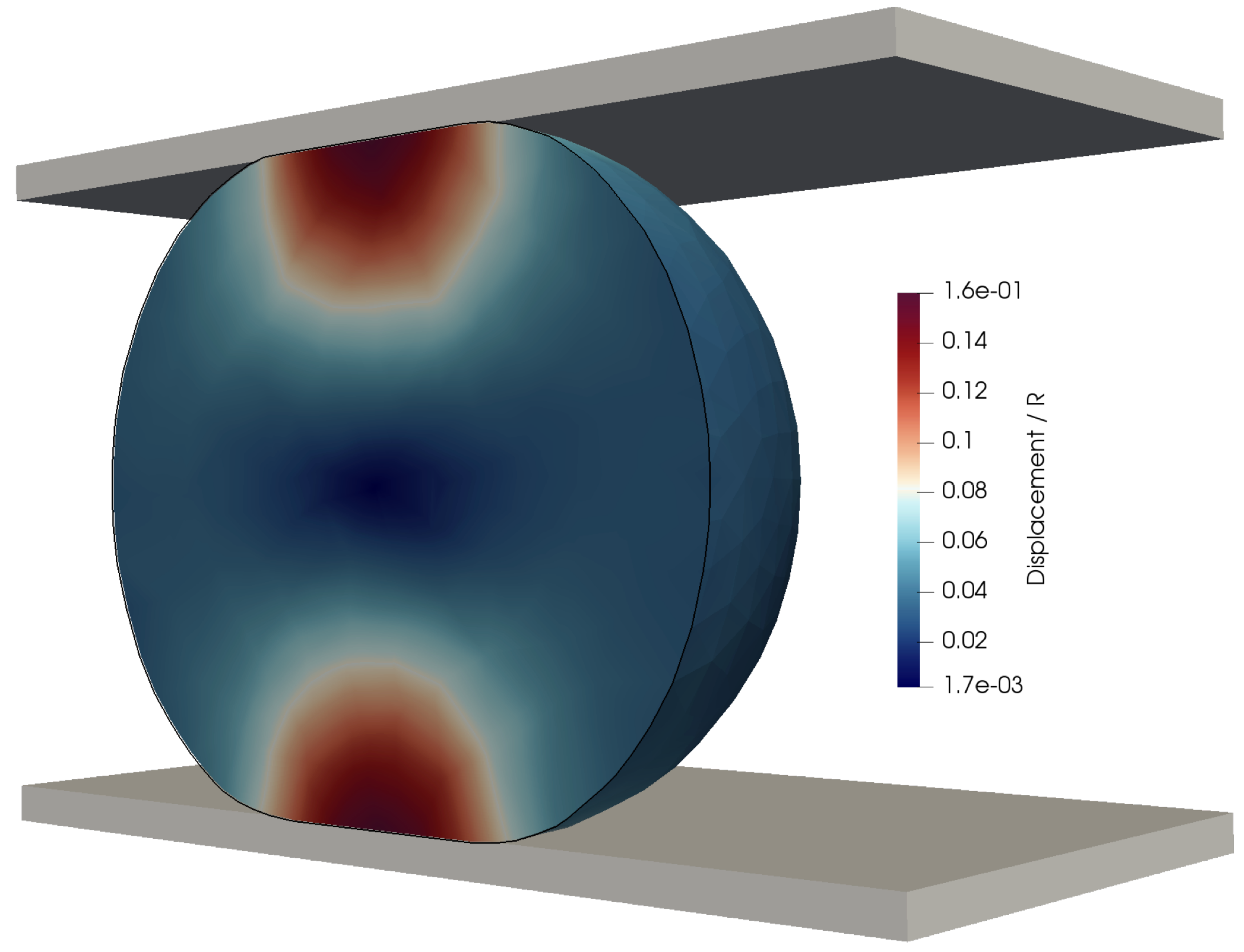} \hspace{20pt} (a)
\includegraphics[width=0.91\linewidth]{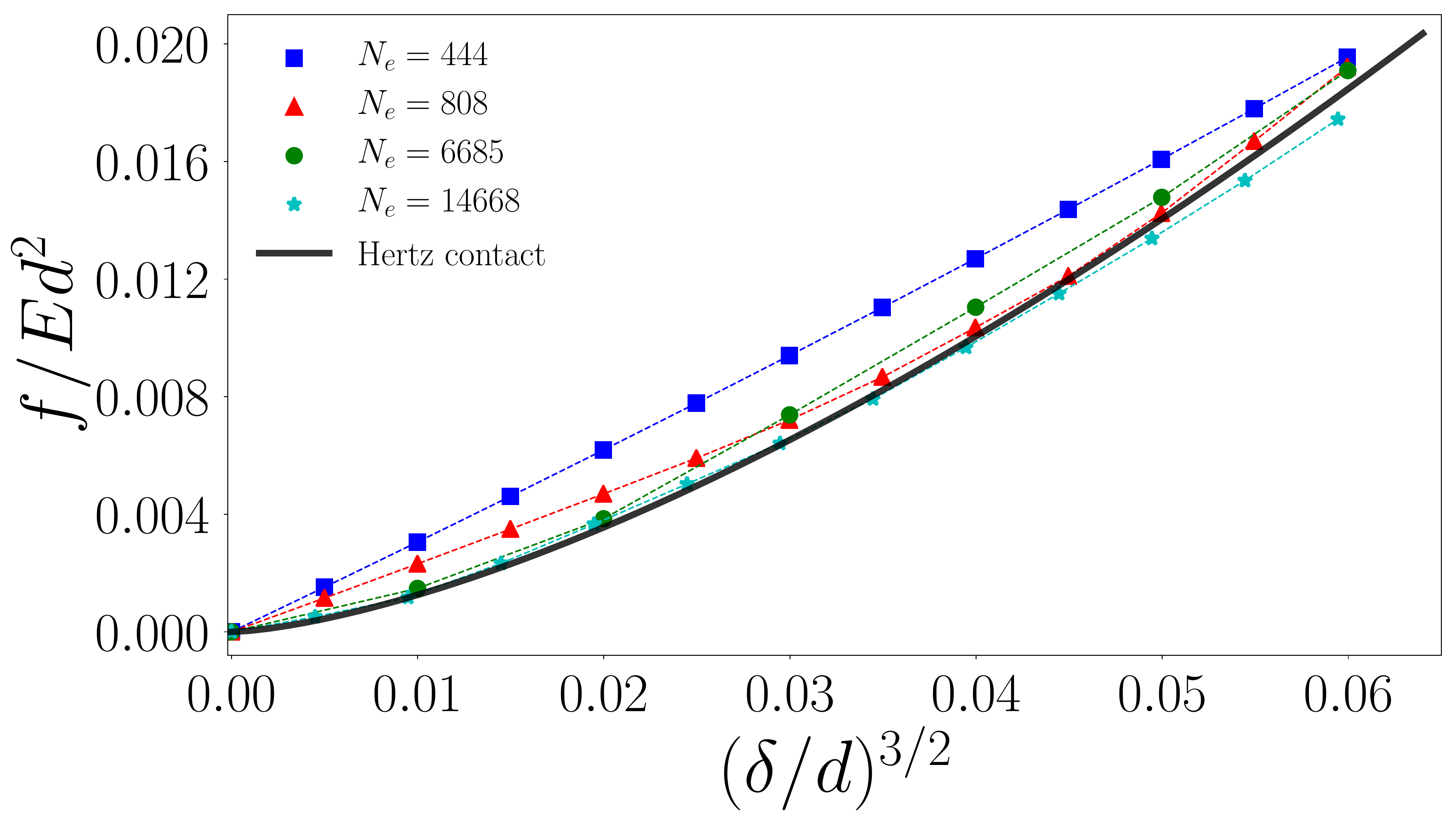}(b)
\caption{(a) 3D cross-section of an elastic spherical particle vertically compressed between two walls. The color intensity is proportional to the mean displacement field.
(b) Normal contact force applied on a single spherical particle as a function of the deformation for different meshes. The continuous black line is the approximation given by the Hertz's contact law Eq. (\ref{Hert_wall_sphere}).}
 \label{Mes_part}
\end{figure}

When dealing with three-dimensional and highly deformable particles, a problematic issue is to find the best compromise between sample representativeness and numerical efficiency.
In this study, we are interested in the isotropic compression of elastic spherical particles.
Therefore, one necessary condition is to verify that the mesh used is, at least, sufficiently accurate concerning the Hertz approximation in the range of small deformations \cite{ johnsonContactMechanics1985}. 
Let us first consider the case of an elastic spherical particle of diameter $d$, with a Poisson's ratio $\nu$ equals $0.495$ and a Young modulus $E$. The sphere is compressed axially as shown in Fig. \ref{Mes_part}(a).
The bottom wall is fixed while the top wall moves downwards at a constant velocity $v_0$ chosen, such as the inertial effects are negligible (i.e., $I<<1$), where $ I=v_0 \sqrt{\rho_0/E}$ \cite{gdrmidiDenseGranularFlows2004}, with $\rho_0$ being the density of the particles.
Figure \ref{Mes_part}(b) shows the evolution of the normal force $f$ as a function of the vertical displacement $\delta$ using $444$, $808$, $6685$ and $14688$ tetrahedral elements with four nodes, together with the corresponding prediction of the Hertz law given by \cite{johnsonContactMechanics1985}:
\begin{equation}
\label{Hert_wall_sphere}
\frac{f}{d^2} = \frac{2^{3/2}}{3} \frac{E}{1-\nu^2} {\left(\frac{\delta}{d}\right)}^{3/2}.
\end{equation}
Compared to this equation, we obtain a good prediction with $14688$ and $6685$ elements, while with $444$ elements, it shows decreasing accuracy.
On the contrary, with $808$ elements, the force-displacement relation is slightly overestimated at the beginning of the deformation, but the response quickly reaches the prediction at higher deformation.
Following this simple analysis, we fixed the number of elements to $808$ for all the simulations presented below.

Concerning the number of particles in an assembly, we adopt a double approach.
First, we rely on previously published works in which it is shown that a number between $32$ and $200$ particles is sufficient to qualitatively represent the loading surfaces, the compaction, or the plastic flow of a compressed assembly of deformable spherical particles \cite{HARTHONG2012784, Shmidt_2010, ABDELMOULA2017142, ZHOU2020153226, PENG2021478}. Second, we use a statistical analysis by considering different samples and focus on their averaged behavior.

Thus, in this study, we consider $8$ systems, $4$ composed of $N=50$ particles and $4$ composed of $N = 100$ particles. For each system, particles are spheres made of an elastic material with Poisson's ratio equals to $0.495$.
The particles are first randomly dropped in a cubic box with a small particle size dispersity around their mean diameter  $\left<d\right>$ in order to avoid crystallization ($ d \in \left[0.8 \left<d \right>,1.2 \left<d \right> \right]$). All packings are then isotropically compressed under a stress $\sigma_0$, such that $\sigma_0/E<<1$ (i.e., the particles can be considered as rigid, in comparison to the applied stress).
This initial compression ends when the change of the packing fraction $\phi$ is below $0.01\%$. After this point, all systems can be considered at the jammed state, characterized by the initial packing fraction $\phi_0$.

\begin{figure}
\centering
\includegraphics[width=1\linewidth]{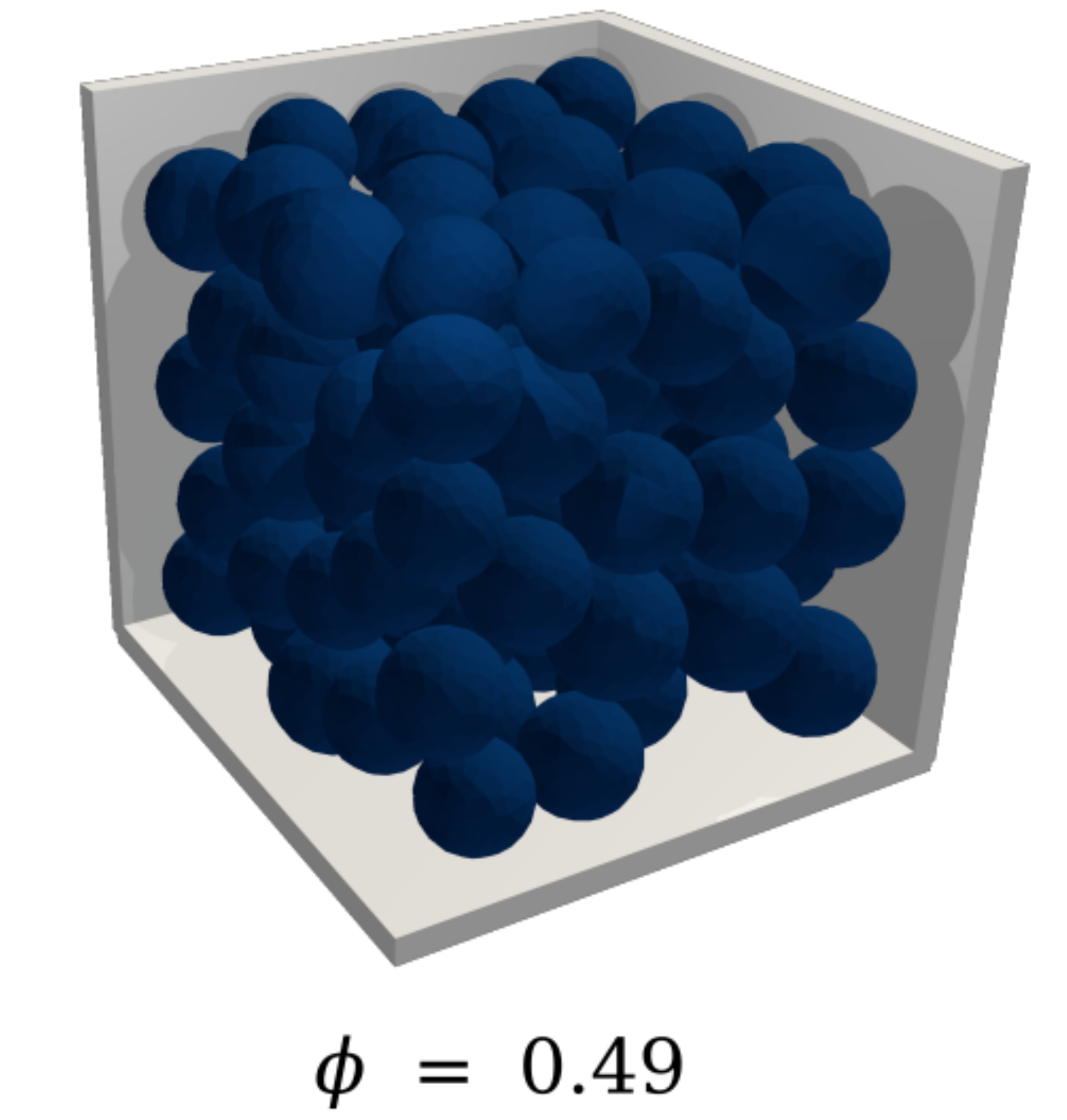}
\includegraphics[width=1\linewidth]{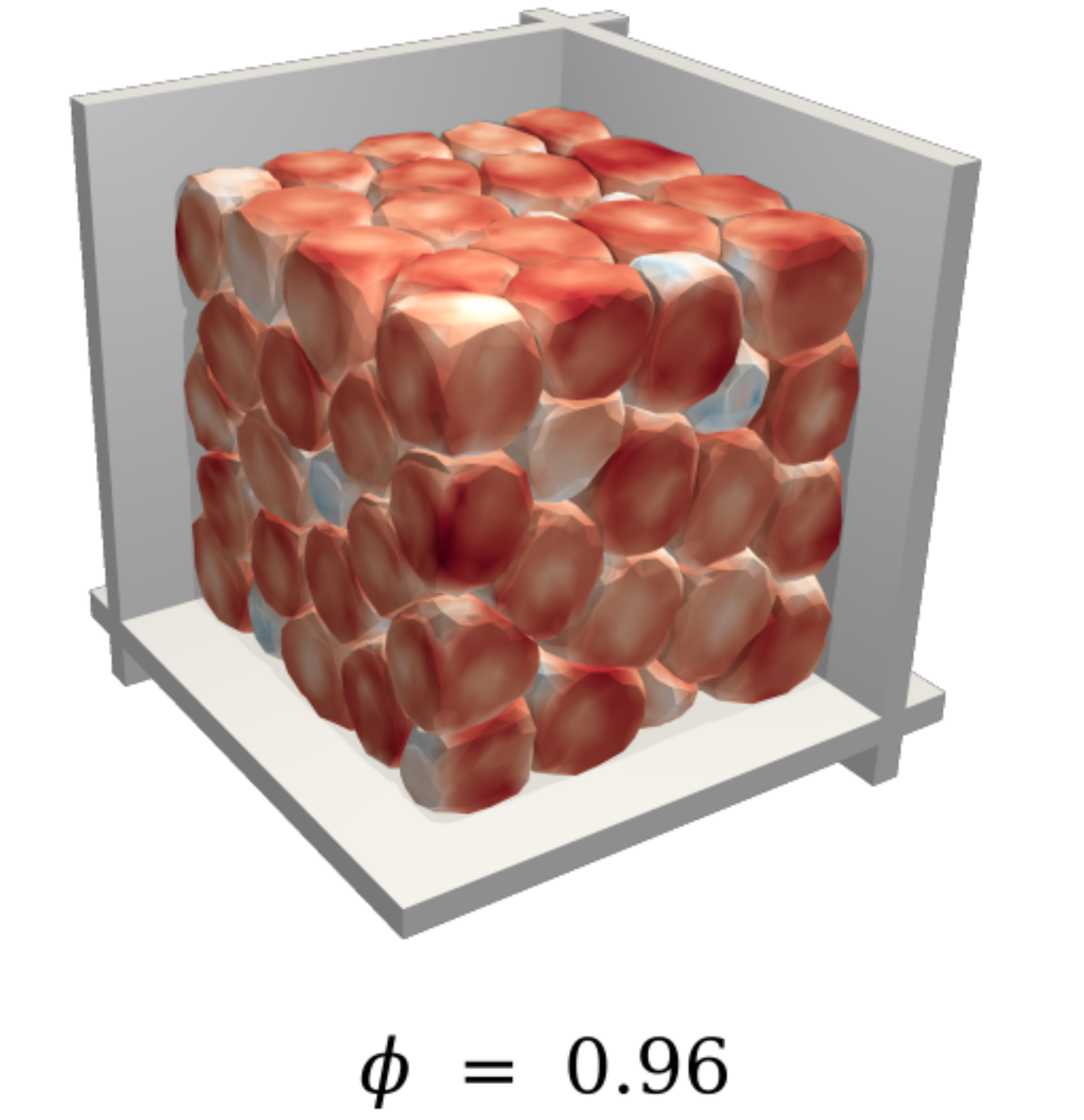}
\caption{View of a granular assembly composed of $100$ soft spherical particles at (a) the initial configuration and (b) close to $\phi=0.96$.
The color intensity, from blue to red, is related to the mean stresses in the particles.}
 \label{Snapshot}
\end{figure}

Then, the packings are isotropically compressed by imposing a constant velocity $v$ on the box's boundaries. The velocity $v$ is carefully chosen to ensure that the systems are always in the quasi-static regime, characterized by an inertial number $I<<1$.
In our simulations, we use a constant friction coefficient between particles $\mu = 0.3$, and we keep the friction coefficient with the walls and gravity equal to zero.
Figure \ref{Snapshot} presents screenshots of an assembly composed of $100$ particles at the jammed state ($\phi\sim0.49$) and near to the maximal dense state with $\phi\sim0.96$.
In the following, the mean behavior for systems composed of $N_p=50$ and $N_p=100$ particles is obtained by averaging over the $4$ corresponding independent sets.
The average jammed packing fraction $\phi_0$ obtained is $0.5$ for $50$ particle assemblies and $0.51$ for $100$ particle assemblies. 

\section{Results}
\label{Sec_results}

\subsection{Packing compaction and particle connectivity}
\label{Stress_evolution}
In this section we analyze the compaction of the assemblies, characterized by the evolution of the packing fraction $\phi$ as a function of the mean confining stress $P$ and the mean particle connectivity $Z$.
The mean confining stress, in the granular system, is extracted from the granular stress tensor $\bm \sigma$, which is computed at each step of the compression as \cite{moreauStressTensorGranular2009}:
\begin{equation}
{\sigma_{\alpha \beta} } = \frac{1}{V} \sum_{c \in V} {f_{\alpha}^c \ell_{\beta}^c},
\label{eq:sigma}
\end{equation}
where $\alpha$ and $\beta$ correspond to $x$, $y$ or $z$, $f_{\alpha}^c$ is the $\alpha^{\rm{th}}$ component of the contact force at the contact $c$, and $\ell_{\beta}^c$ is the $\beta^{\rm{th}}$ component of the vector that join the two centers of the particles interacting at the contact $c$.
Note that the total contact force between two deformable particles is computed as the vectorial sum of the forces at the contact nodes along the shared interface.
The mean confining stress is then given by $P = (\sigma_{1} + \sigma_{2} + \sigma_{3})/3$, where $\sigma_{1}$, $\sigma_{2}$ and $\sigma_{3}$ are the principal stress values
of $\bm \sigma$. The packing fraction $\phi$ is also related to the macroscopic deformation $\epsilon$, by $\epsilon = - \ln(\phi_0/\phi)$.

\begin{figure}
\centering
\includegraphics[width=0.8\linewidth]{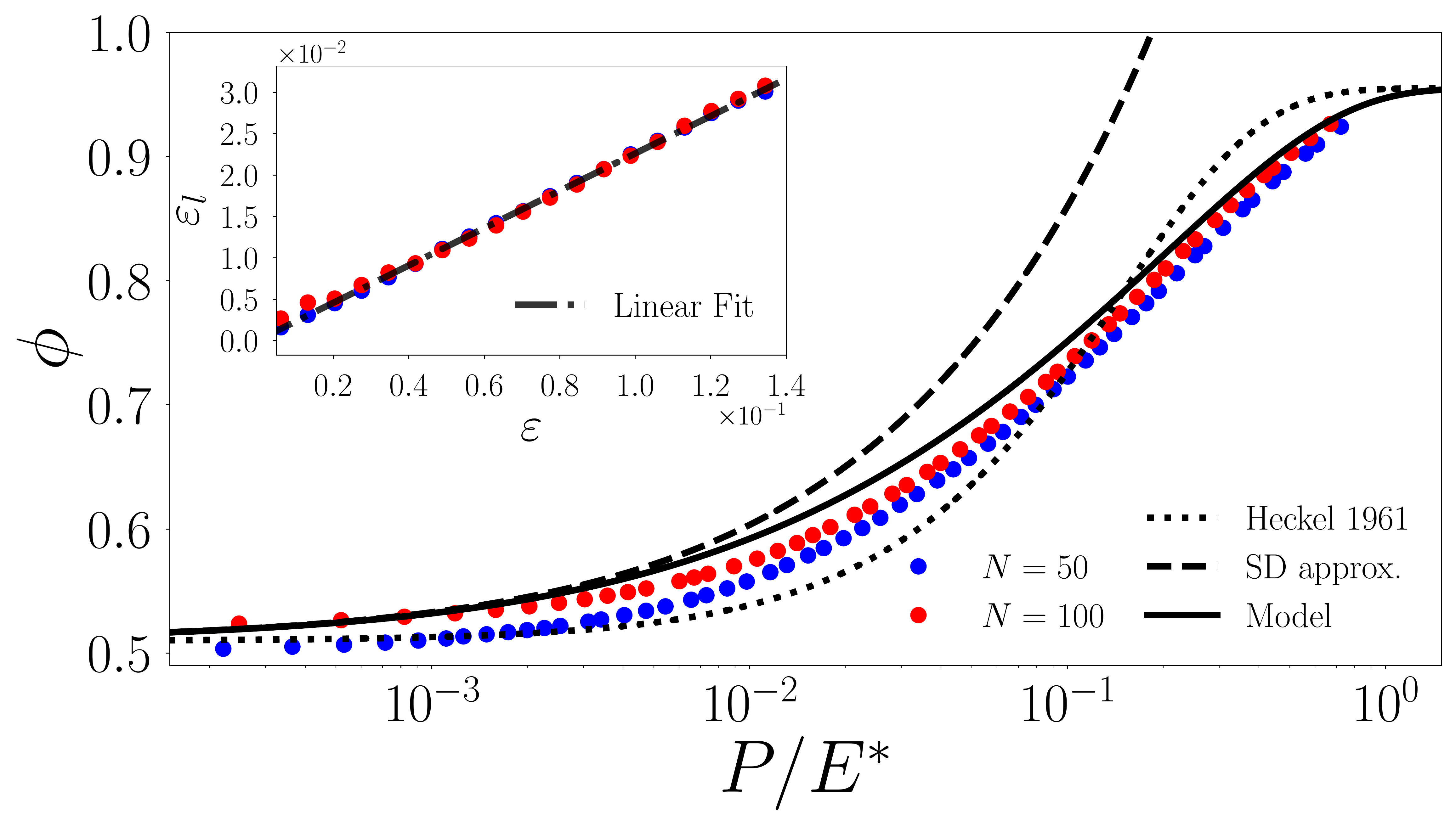}(a)
\includegraphics[width=0.8\linewidth]{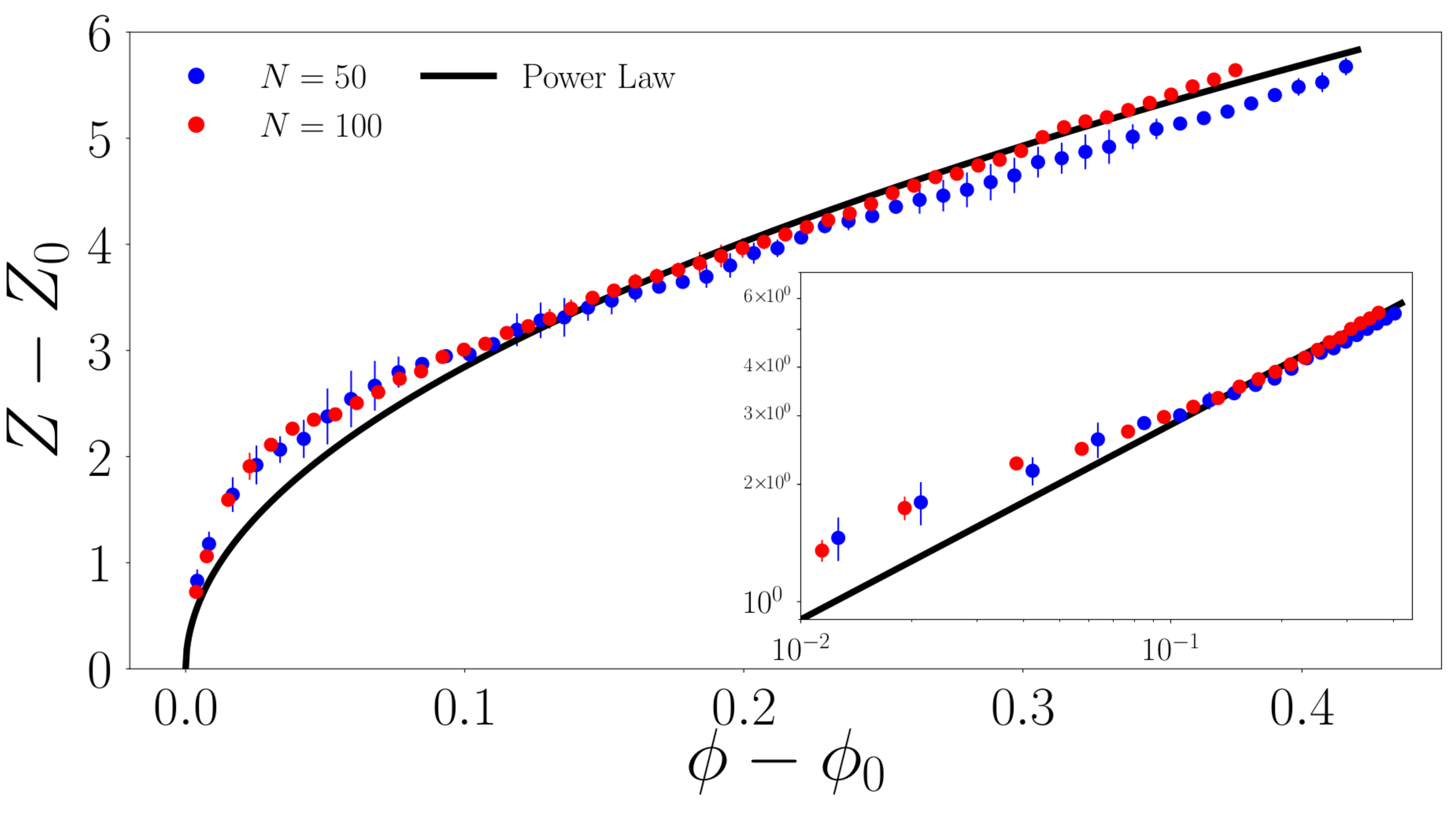}(b)
\caption{(a) Packing fraction $\phi$ as a function of the mean confining stress $P$ normalized by the reduced Young Modulus $E^*$.
The dotted line is the approximation given by Heckel Eq. (\ref{Eq_Heckel}), the dashed line is the small strain approximation given by Eq. (\ref{eq:Pgsd}) (SD), and the continuous black line is the prediction given by our micromechanical approach Eq. (\ref{eq:Pglobal_2}).
The inset shows the macroscopic volumetric strain $\varepsilon$ as a function of the mean contact strain $\langle \epsilon_\ell \rangle$ in the small deformation domain.
(b) Reduced coordination number $Z - Z_0$ as a function of the reduced solid fraction $\phi-\phi_0$ (log-log scale is shown in the inset).
The continuous black line is the power-law relation given by Eq. (\ref{eq:ZvsPpi}) with exponent $0.5$.
Error bars represent the standard deviation on the averaged behavior performed over $4$ independent samples.}
 \label{PPhi_et_ZPhi}
\end{figure}

Figure \ref{PPhi_et_ZPhi}(a) shows the evolution of $\phi$ as a function of the mean confining stress $P$, normalized by the reduced Young Modulus $E^*=E/2(1-\nu^2)$, for the assemblies composed of $50$ and $100$ particles.
From the jammed state, the packing fraction asymptotically increases towards the value $\phi_{max}$, at high pressure.

Note that the compaction curves for the $50$ and $100$ particle systems collapse on the same curve, which is in agreement with the previous works mentioning the minimum number of grains necessary to capture the average comparative behavior. On these compaction curves, we also show the approximation proposed by Heckel-Secondi, with the following form \cite{Heckel1961,Secondi2002_Modelling}:
\begin{equation}
\frac{P}{E^*} = -A \ln\left( \frac{\phi_{max}-\phi}{\phi_{max}-\phi_0}  \right),
\label{Eq_Heckel}
\end{equation}
with $A$ a fitting constant equal to $0.15$ in our case, and $\phi_{max}=0.965$.

Equation (\ref{Eq_Heckel}), although very simple in its form, is able to capture the general tendency of the compaction but slightly mismatches its evolution for intermediate pressures.
Also, the parameter $A$ does not have a well-established physical meaning, and different values may be required to fit the data depending on the friction coefficient \cite{Cardenas2020_Compaction}or the bulk behavior of the particles \cite{Carroll1984, Platzer2018}.
Some improvements to the Heckel-Secondi equation have been proposed by Ge et al. \cite{Ge1995_A}, Zhang et al. \cite{Zhang2014}, and Wunsch et al. \cite{Wunsch2019_A} by considering a double log approach (i.e., $\ln P \propto \log \ln \phi$). However, unlike the Heckel-Secondi equation, which can be justified \cite{Carroll1984}, these new approaches rely only on data fitting. 

In Fig.\ref{PPhi_et_ZPhi}(b), we plot the evolution of the mean particle connectivity $Z$ as a function of $\phi$.
At the jammed state, the packing structure is characterized by a minimal value $Z_0$, which depends on the coefficient of friction, the packing preparation, and the shape of the particles \cite{Hecke2009_Jamming, Donev2005_Pair, smithAthermalJammingSoft2010}. For spherical assemblies, $Z_0$ is equal to $6$ when the friction vanishes, and it varies between $4$ and $6$ for higher friction coefficients.
In our two frictional systems, we find $Z_0\simeq 3.5$.
Beyond the jammed state, $Z$ continues to increase and, as shown in the previous 2D numerical \cite{Vu2020_compaction, Cardenas2020_Compaction, Cardenas_pentagons_2021} and experimental studies \cite{Majmudar2007_Jamming, Katgert2010_Jamming, Vu2019}, this increase follows a power law with exponent $1/2$:
\begin{equation}\label{eq:ZvsPpi}
(Z - Z_0) = \psi \sqrt{\phi - \phi_0},
\end{equation}
with $\psi \approx 8.5$, a constant fully defined through the characteristics of the jammed state and the final dense state as $\psi = (Z_{max}-Z_0)/\sqrt{\phi_{max}-\phi_0}$, with
$Z_{max}$ the maximum packing fraction as $\phi \rightarrow\phi_{max}$.
Thus, this power-law relation already observed in 2D can now be extended to the case of 3D soft particle assemblies.

\subsection{A compaction equation}
\label{Theory}
As discussed in the introduction, there are many compaction equations trying to relate the confining stress to the evolution of the packing fraction.
Some of them, such as the Heckel-Secondi equation, although based on fitting parameters, can be justified by macroscopic arguments \cite{Carroll1984, Kim1987}.
However, the vast majority are settled on adjustment strategies sometimes involving several fitting parameters.
In this section, we briefly recall the general framework introduced in a previous study \cite{Cardenas_pentagons_2021} allowing us to relate the packing fraction to the applied stress through the micromechanical specificity of a given system. We then apply this general framework to the case of three-dimensional soft spherical particles.

The stress tensor Eq. (\ref{eq:sigma}) can be rewritten, as a sum over all contacts, as:
\begin{equation}
\sigma_{\alpha \beta} = n_c \langle f^c_{\alpha}\ell^c_{\beta} \rangle_c,
\label{eq:sigma_contact}
\end{equation}
where $\langle...\rangle_c$ is the average over all contacts. The density of contacts $n_c$ is given by $n_c=N_c/V$, with $N_c$ the total number of contacts in the volume $V$.
Considering a small particle size distribution around the diameter $\langle d \rangle$, $\sum_{p\in V} V_p\simeq N_pV_p$, with $V_p = (\pi/6)d^3$, the contact density can be rewritten as $n_c \simeq 3Z\phi/(\pi d^3)$, with $Z=2N_c/N_p$.
From the definition of $P$ via the principal stresses of $\sigma$, we get \cite{Rothenburg1989_Analytical, Agnolin2007c, Agnolin2008_On, Khalili2017b}:
\begin{equation}\label{eq:Pglobal_local_contact}
P \simeq \frac{ \phi Z} {\pi} \sigma_{\ell},
\end{equation}
with $\sigma_{\ell} = \langle f^c \cdot \ell^c \rangle_c/\langle d \rangle^3$, a measure of the mean contact stress.
This way of writing $P$ as a function of $Z$, $\phi$, and $\sigma_\ell$ is, in fact, very common and has been successfully applied in different contexts.
For example, it has been used to relate the bulk properties of an assembly to the elastic contact properties \cite{Brodu2015b,Khalili2017a}, or to link the macroscopic cohesive strength to the cohesive behavior between the interface of particles in contact \cite{Richefeu2006, Azema2018}.
The Equation (\ref{eq:Pglobal_local_contact}) reveals the active role of the evolution of the microstructure in the evolution of $P$ with $\phi$.

First, we focus on the small deformation domain. We can rely on Hertz's prediction where, in 3D, the force $f^c$ at a contact $c$ between two touching particles is related to the contact deflection $\delta^c$ by $f^c = (2/3) E^* d^{1/2} {\delta^c}^{3/2}$.
Then, since $\ell_c\sim d$ in the range of small deformations, we get $\sigma_\ell \sim (2/3)E^* \langle \varepsilon_{\ell} \rangle^{3/2}$, where $\varepsilon_{\ell} = \delta^c/d$ is the deformation at a contact $c$, assuming that $\langle \varepsilon_{\ell}^{3/2}\rangle \sim \langle \varepsilon_{\ell} \rangle^{3/2}$ which is well verified in our weakly polydisperse systems. Also, with a good approximation, we get that $Z=Z_0$.
Finally, our simulations show that the mean contact strain $\langle \epsilon_\ell \rangle$ and
the macroscopic volumetric strain are linearly dependent as $\langle \epsilon_\ell \rangle \sim (1/\Gamma) \varepsilon$, with $\Gamma\sim4.4$ (see inset in Fig.\ref{PPhi_et_ZPhi}(a)).
This value is close to the one obtained in 2D with disks and non-circular particles \cite{Cardenas_pentagons_2021}. Note that $\Gamma=3$ in the ideal case of a cubic lattice arrangement of spheres.
Finally, by considering all these ingredients, Eq. (\ref{eq:Pglobal_local_contact}) is rewritten as:
\begin{equation}
\label{eq:Pgsd}
\frac{P_{SD}}{E^*} = -\frac{2}{3\pi\Gamma^{3/2}}Z_0\phi \ln^{(3/2)}\left(\frac{\phi_0}{\phi}\right),
\end{equation}
with $P_{SD}$ the limit of $P(\phi)$ at small deformations.
The prediction given by Eq. (\ref{eq:Pgsd}) is shown in Fig. \ref{PPhi_et_ZPhi}.
As expected, we see a fair approximation of the compaction evolution in the small-strain domain, but it fails to predict the evolution at larger strains.

The critical issue for the large strain domain is to find a proper approximation of $\sigma_\ell(\phi)$.
To find so, we can combine the previous microscopic approach with a macroscopic development by Carroll and Kim \cite{Carroll1984, Kim1987}.
Assuming that the compaction behavior can be equivalent to the collapse of a cavity within the elastic medium, they showed that $P \propto \ln[(\phi_{max}-\phi)/(\phi_{max}-\phi_0)]$.
Using this macroscopic approximation together with the micromechanical expression of $P$ given by Eq. (\ref{eq:Pglobal_local_contact}), and remarking  that the quantity $Z\phi$ is finite, it is easy to show that, necessarily, $\sigma_\ell = \alpha(\phi)  \ln[(\phi_{max}-\phi)/(\phi_{max}-\phi_0)]$, with $\alpha$ a function that depends, {\it a priori}, on $\phi$.
Then, by ($i$) introducing the above form of $\sigma_\ell$ into Eq. (\ref{eq:Pglobal_local_contact}), ($ii$) ensuring the continuity to small deformation (i.e., $P\rightarrow P_{SD}$ for $\phi\rightarrow \phi_0$), and ($iii$) introducing the $Z-\phi$ relation (Eq. (\ref{eq:ZvsPpi})) into Eq. (\ref{eq:Pglobal_local_contact}), we get:
\begin{equation}\label{eq:Pglobal_2}
\frac{P}{E^*} = -\frac{2}{3\pi\Gamma^{3/2}}\left(\frac{\phi_{max}-\phi_0}{\phi_0^{3/2}}\right) \phi  \sqrt{\phi - \phi_0}  \left[Z_0 - \psi \sqrt{\phi-\phi_0}\right] \ln\left( \frac{\phi_{max}-\phi}{\phi_{max}-\phi_0} \right).
\end{equation}

The compaction equation given by Eq. (\ref{eq:Pglobal_2}) is plotted with a black continuous line in Fig. \ref{PPhi_et_ZPhi}(a) together with our numerical data.
The prediction is able to capture the asymptotic behavior close to the jammed state and the asymptotic behavior at high pressures.
In Eq. (\ref{eq:Pglobal_2}), and in contrast to previous models, only one parameter, the maximum packing fraction $\phi_{max}$, is unknown. Other constants are entirely determined through the initial jammed state and the mapping between the packing fraction and coordination curve.

Finally, it is worth mentioning the differences between Eq. (\ref{eq:Pglobal_2}) and its two-dimensional equivalent \cite{Cantor2020_Compaction,Cardenas_pentagons_2021}.
In two dimensions, the numerical simulations show that $\sigma_\ell$ depends linearly on the mean contact strain, consistently with the approximation classically done in 2D MD-like simulations \cite{cundall1979}. This linear dependence in 2D then simplifies the development by replacing the terms $(2\sqrt{\phi - \phi_0})/(3\Gamma\phi_0)^{3/2})$ in the Eq. (\ref{eq:Pglobal_2}) by only $1/(\Gamma\phi_0)$.

\subsection{Particle shape and particle stress distribution}
\label{Local_stress}
During the compression, the shape of the particles evolves from an initial spherical shape to a polyhedral shape, which also modifies the stress distribution.

\begin{figure}
\centering
\includegraphics[width=0.9\linewidth]{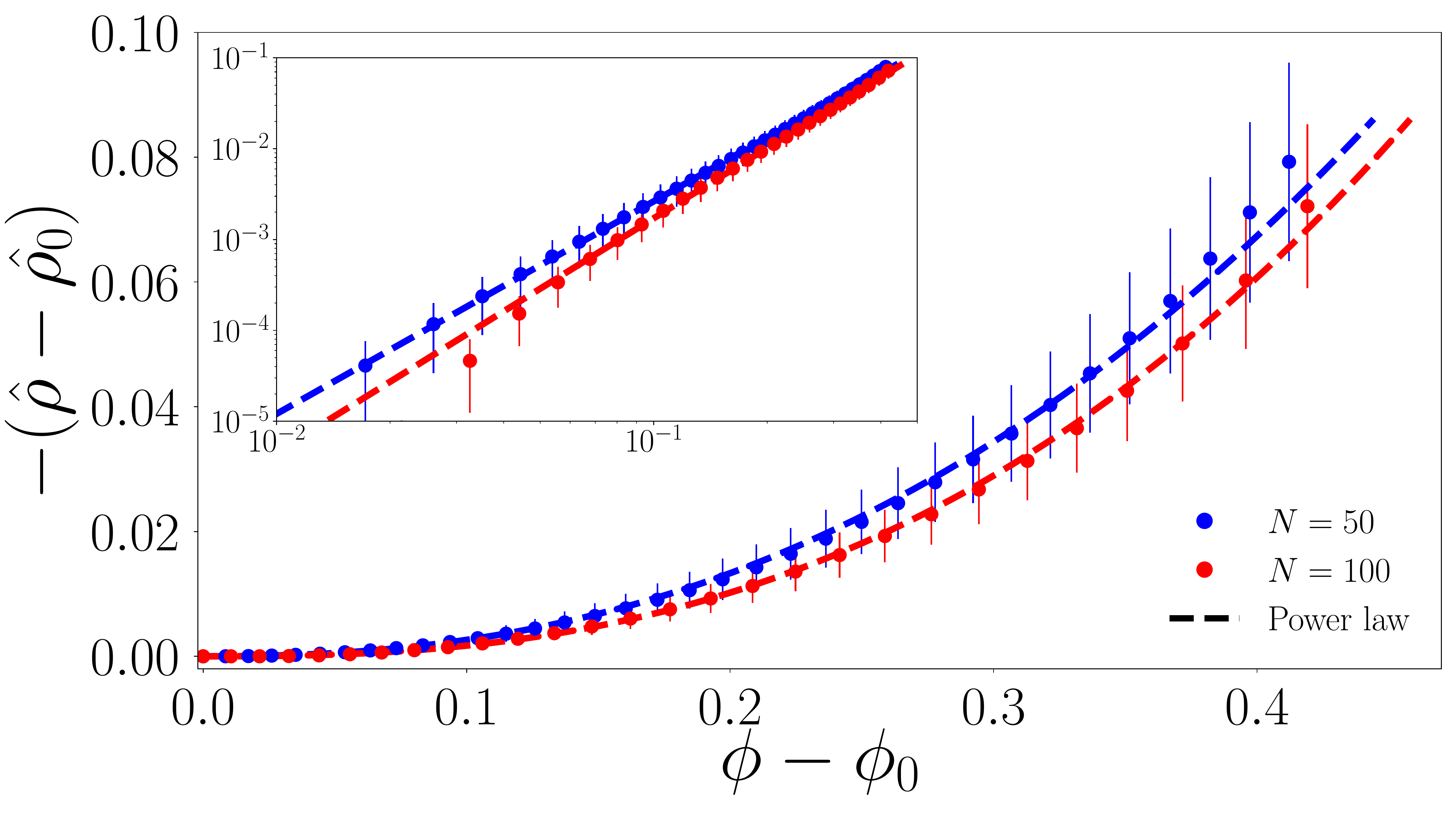}
\caption{Evolution of the excess sphericity, $\hat{\rho}-\hat{\rho}_0$ as a function of the excess packing fraction $\phi-\phi_0$ for the isotropic compaction of soft spheres. The dashed line is the power-law relation given by Eq. (\ref{Eq_shape_phi_3D}).
}
 \label{fig:shape_3D}
\end{figure}

At the lowest order, the shape of the particles can be characterized by means of the sphericity parameter $\hat{\rho}$, defined by:
\begin{equation} \label{eq:circu_mix}
\hat{\rho} = \left\langle \pi^{1/3} \frac{(6V_i)^{2/3}}{a_i} \right\rangle_i,
\end{equation}
with $a_i$ the surface area of the particle and $\langle ...\rangle_i$ the average over the particles in the volume $V$. By definition, the sphericity of a sphere is one, with values below one for any other geometry.
In Fig. \ref{fig:shape_3D}, we plot the evolution of $(\hat{\rho}-\hat{\rho}_0)$, with $\hat{\rho}_0\sim 1$ the initial sphericity of the particles, as a function of the excess packing fraction, $\phi-\phi_0$.
We find that the shape parameter increases as a power law with exponent $\beta$:
\begin{equation}
 \label{Eq_shape_phi_3D}
 \hat{\rho}-\hat{\rho}_0 = A (\phi-\phi_0)^{\beta},
\end{equation}
with $\beta \approx 2.5$ and $A \approx 0.6$. 
It is interesting to note that a similar tendency has been recently observed in 2D for soft-disks assemblies \cite{} with a similar exponent, which evidences a seemingly universal geometrical characteristic of the compaction of rounded soft particles, as for the relation between $Z$ and $\phi$.

\begin{figure*}
\centering
\includegraphics[width=0.28\linewidth]{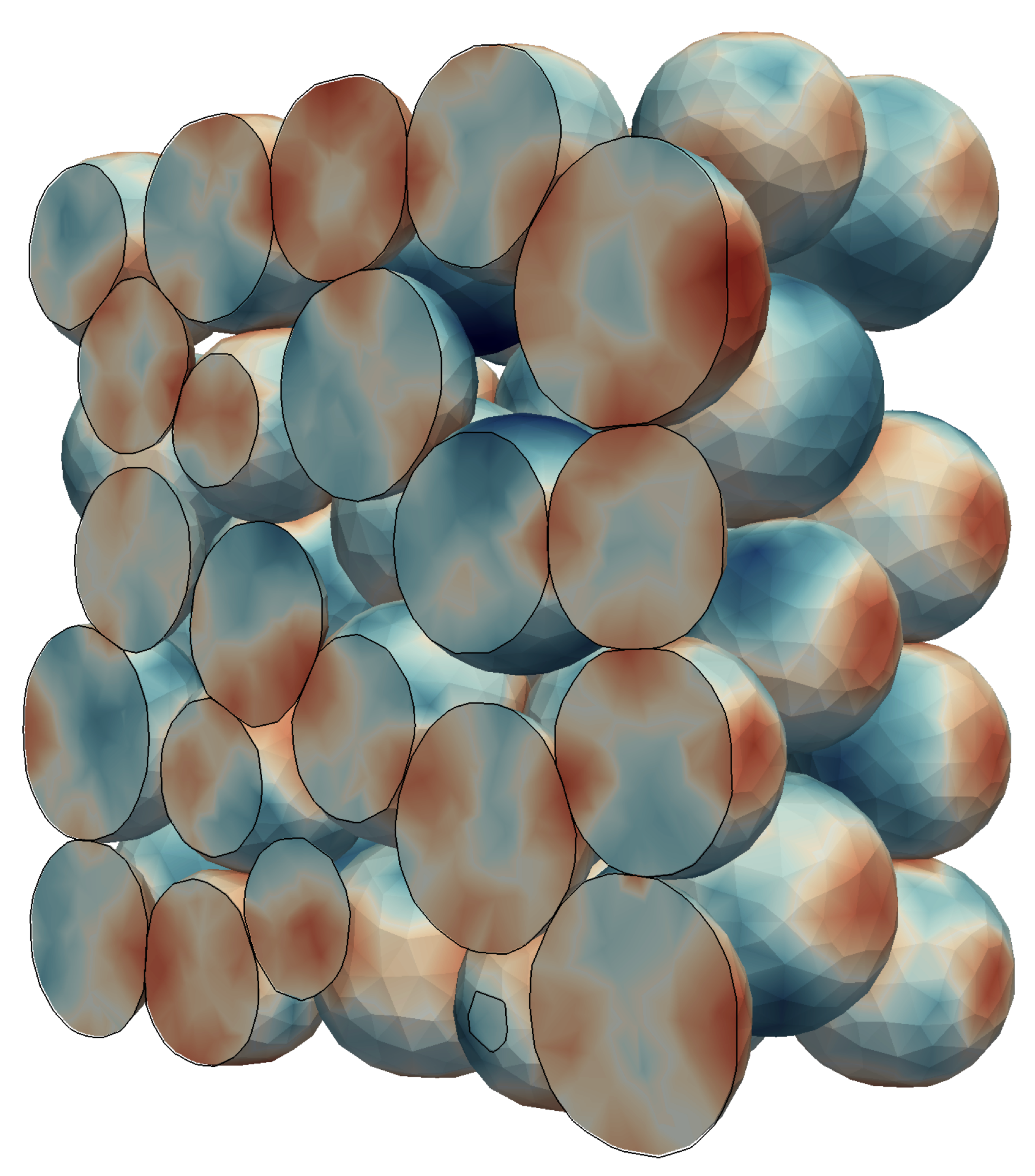}(a)
\includegraphics[width=0.28\linewidth]{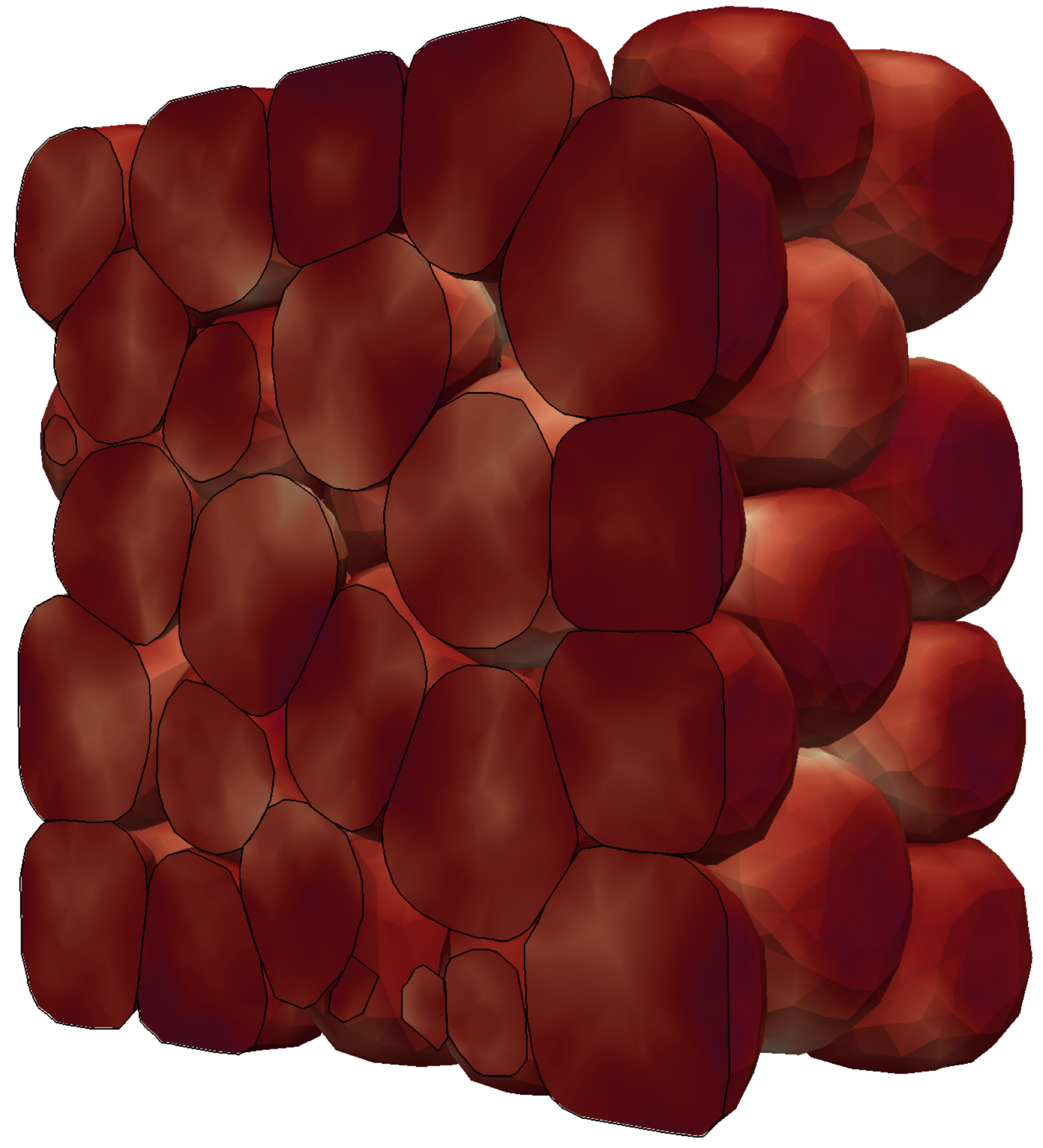}(b)
\includegraphics[width=0.28\linewidth]{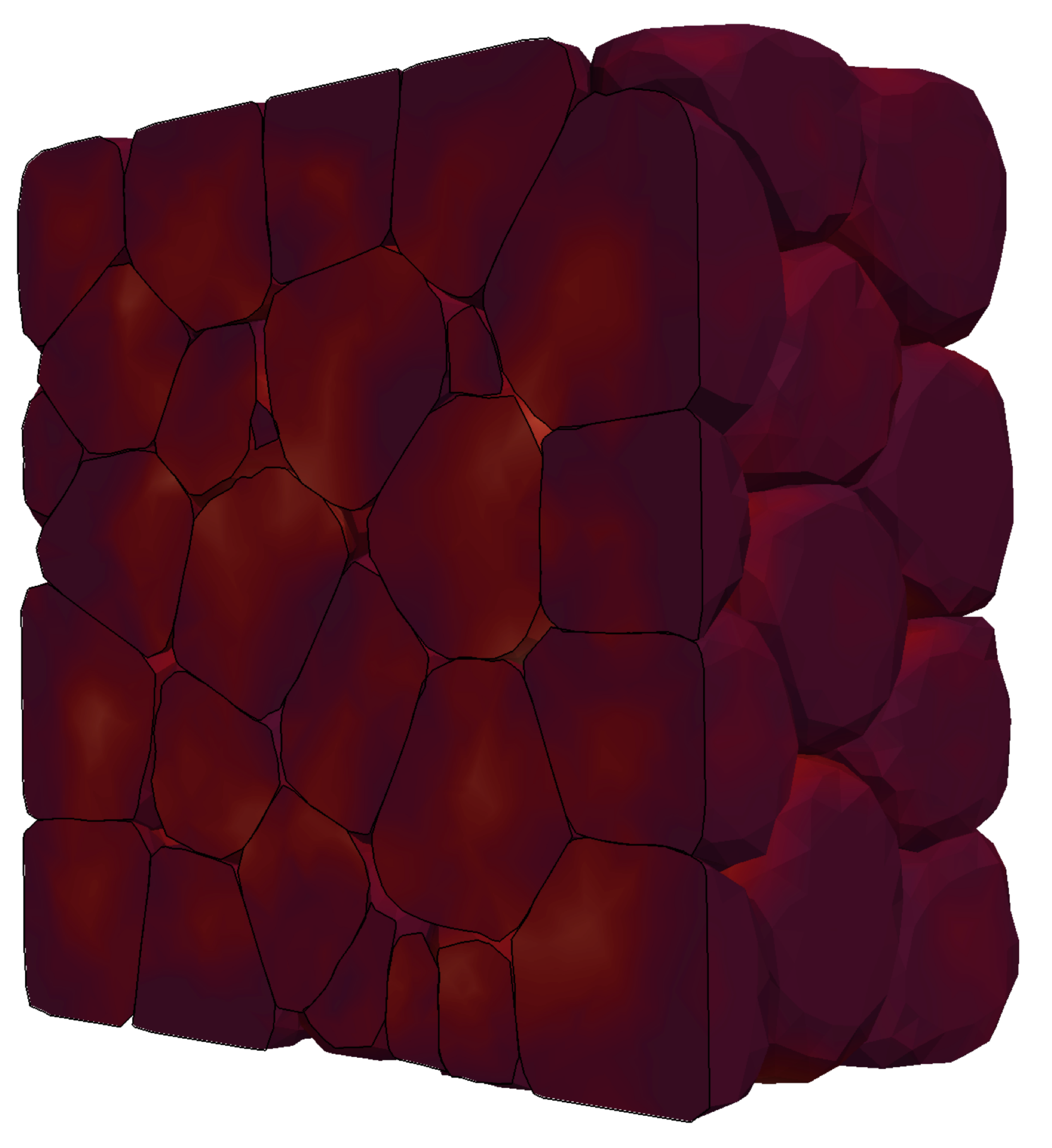}(c)
\includegraphics[width=0.25\linewidth]{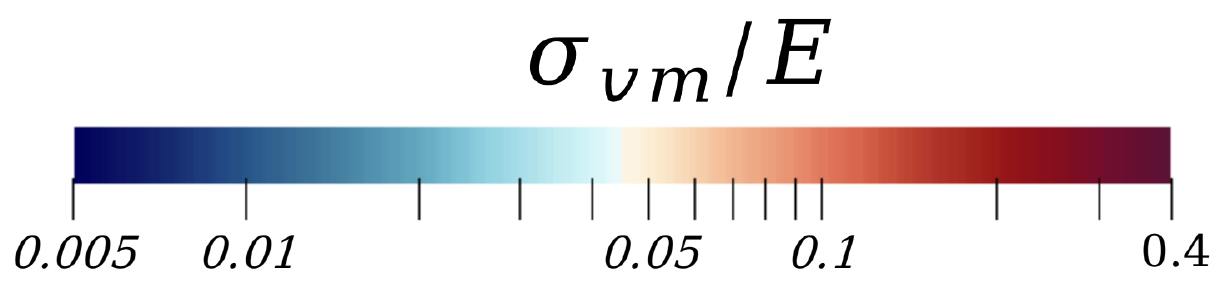}
\caption{Three dimensional cross-section of the von Misses stress field $\sigma_{vm}$ at each particle and for different packing fraction $\phi=0.66$ (a), $\phi=0.86$ (b) and $\phi=0.96$ (c) in an assembly of 100 particles.
The color intensity is proportional to the von Mises stress scaled by the Young modulus, $\sigma_{vm}/E$.}
 \label{snap_pe}
\end{figure*}

The change in grain shape is necessarily coupled with a redistribution of stresses within the grains. Thus, let us consider the Cauchy stress tensor $\boldsymbol \sigma^{{C}}$ calculated inside the grains.
Note that $\boldsymbol \sigma^{{C}}$ should not be confused with the granular stress tensor ${\boldsymbol \sigma}$ defined above and calculated from the contact forces.
Figure \ref{snap_pe} shows a cross-section images of an assembly of $100$ particles, where the color scale represents the von Mises stress computed at each node.
After the jammed state, strong heterogeneities in the stress distribution inside the particles can be seen (see Fig.\ref{snap_pe}(a)).
The grains are mainly deformed at the contact points, which generally support the maximum stress.
Far beyond the jammed state (Fig.\ref{snap_pe}(b,c)), the shape of the grains strongly changes, the size of the pore declines, and the spatial stress distributions tend to homogenize.

\begin{figure}
\centering
\includegraphics[width=0.98\linewidth]{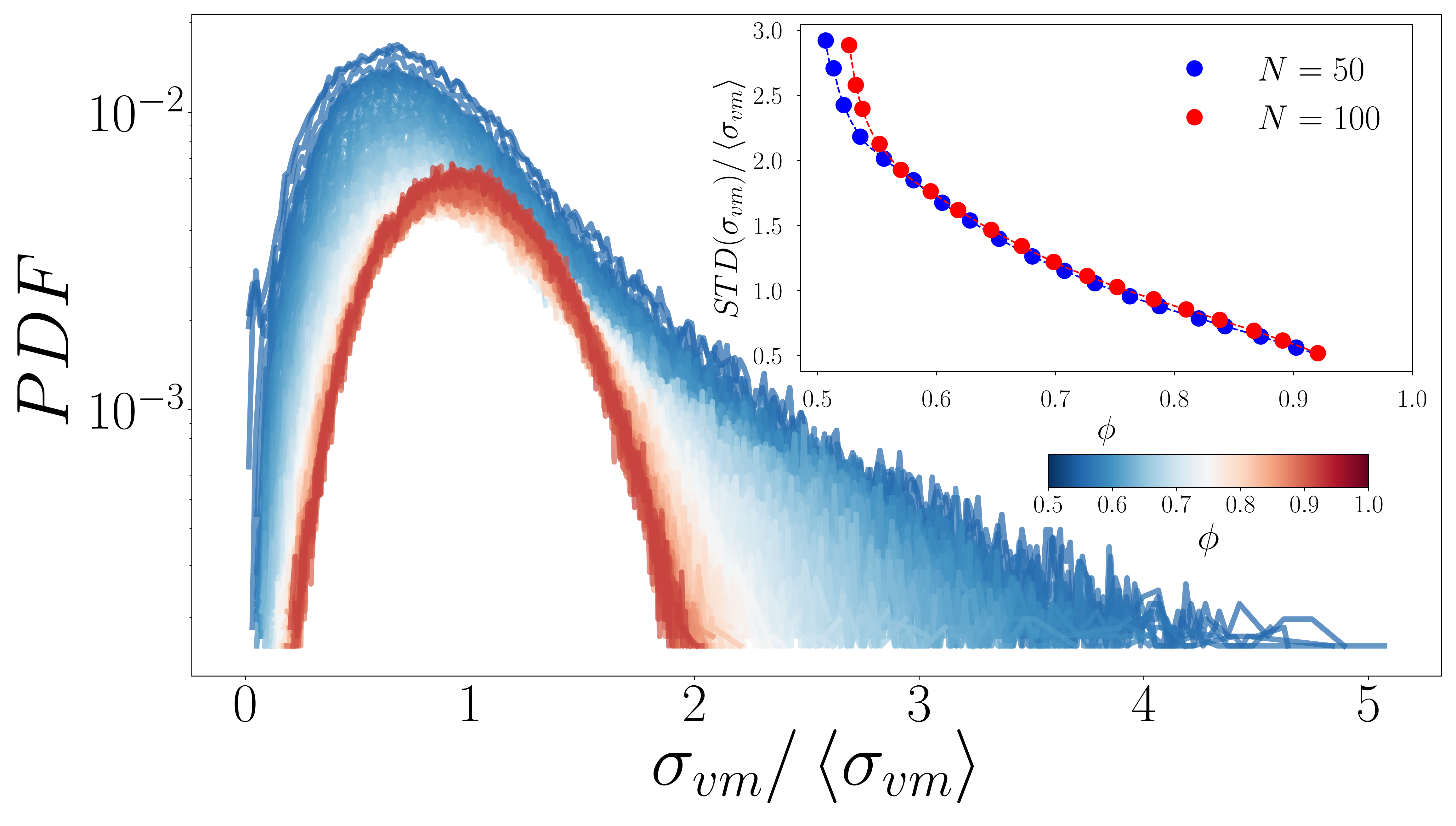}
\caption{Probability density function (PDF) of 
the local von Mises stress $\sigma_{vm}$ computed on each node and normalized by the corresponding average value in one of the systems composed of 100 particles.
The inset shows the standard deviation of the distribution of $\sigma_{vm}$ as a function of the packing fraction $\phi$ for $50$ and $100$ particles assemblies
(averaged over the four independent systems). }
 \label{pdf_pe}
\end{figure}

Fig. \ref{pdf_pe} shows the evolution of the probability density functions (PDF) of the equivalent von Mises stress $\sigma_{vm}$.
Close to the jammed state, we observe exponential decays reminiscent of the distribution of contact forces classically observed in rigid particle assemblies \cite{Nguyen2014_Effect,Mueth1998_Force,Daniels2017_Photoelastic,Abed2019_Enlightening}. This underlines the fact that, although the assembly is isotropically compressed at the macroscopic scale, the particles may undergo large shear stress.
As the packing fraction increases, the PDFs get narrower and gradually transform into Gaussian-like distributions centered around a given mean value.

From these observations, and consistently to the previous observations made in two dimensions, a schematic picture emerges to describe the compaction from a local perspective.
During the compaction, the assembly shifts from a rigid granular material to a continuous-like material.
In the granular-material state, the voids are filled by affine displacement of the particles and small deformations that do not change the spherical shape of the particles significantly.
Then, stress and contact force homogenize within the packing due to the increasing average contact surface and the mean coordination number. This progressive shift of the distributions to Gaussian-like distributions evidence that the system is turning into a more continuous-like material as the packing fraction approaches its maximum value.
This is verified by the decreasing standard deviation of such distributions (inset in Fig. \ref{pdf_pe}(b)).

\section{Conclusions}
\label{Conclu_section}
This paper investigates the compaction behavior of three-dimensional soft spherical particle assemblies through the Non-Smooth Contact Dynamic Method.
From the jammed state to a packing fraction close to $1$, various packings composed of $50$ and $100$ meshed spherical particles were isotropically compressed by applying a constant inward velocity on the boundaries. The mean compaction behavior was analyzed by averaging over the independent initial states.

One of the main results of this work is the writing of a new equation for the compaction of 3D soft spherical particle assemblies based on micromechanical considerations and entirely determined from the structural properties of the packing.
More precisely, this equation is derived from the micromechanical expression of the granular stress tensor together with the approximation of the Hertz contact law between two spherical particles at small strain and assuming a logarithmic shape of the compaction curve at large strain.
Moreover, our numerical data shows that the power-law relation between the coordination number and the packing fraction, after the jamming, is still valid in three-dimensional compaction of elastic spheres, which allows us, {\it in fine}, to write a compaction equation nicely fitting our numerical data. Further, we show that the stress distribution within the particles becomes more homogenous as the packing fraction increases.
Close to the jammed state, the probability density functions of the von Mises stress decrease exponentially as the maximum stress increases.
The distributions progressively shift into a Gaussian-like shape at high packing fractions, which means that the system turns into a more continuous-like material.

The general methodology used for the build-up of this 3D compaction equation was previously implemented in two-dimensional geometries.
Although the compaction curves in two and three dimensions appear to be similar in their overall shape (i.e., in both cases, the packing fraction increases and tends asymptotically to a maximum value as the confining stress increases), it is interesting to note that the equation underlying the variation of $P$ with $\phi$ established with the same micromechanical framework depends on the dimensionality.
The origin of this dependence on the space dimension lies in the functional form of the contact law in the small deformation regime.
Thus, for pursuing a more general compaction equation, it is possible to apply the same micromechanical framework described in this article to assemblies whose particles have more complex behaviors, such as plastic, elastoplastic, or visco-elastoplastic, and also to polydisperse systems.
It will be enough to identify the force law between two particles and integrate it into the framework presented here for all these cases.

Finally, we would also like to point out that, from our best knowledge, this is the first time that the Non-Smooth Contact Dynamics Method is applied to the case of compaction of deformable grains assembly in three dimensions. 

From a purely numerical perspective, many efforts still need to be made on numerical optimization and parallelization of algorithms to increase performance and system size.
In particular, it would be interesting to consider periodic conditions in 3D, at least in two directions. 

We warmly thank Frederic Dubois for the valuable technical advice on the simulations in LMGC90 and the fruitful discussions regarding the numerical strategies for modeling highly deformable particles in the frame of the Non-Smooth Contact Dynamic method, specifically in three dimensions. We also acknowledge the support of the High-Performance Computing Platform MESO@LR.

\section*{Conflicts of interest}
There are no conflicts to declare.
\appendix

\balance

\bibliography{biblio} 
\bibliographystyle{rsc} 

\end{document}